\documentclass[conference, 10pt]{IEEEtran}

\IEEEoverridecommandlockouts

\usepackage{amsmath}    
\usepackage{graphicx}   
\usepackage{epstopdf}
\usepackage{verbatim}   
\usepackage{color}      
\usepackage{subfigure}  
\usepackage{algorithm}
\usepackage{array}
\usepackage{algorithmic}
\usepackage{nomencl}
\usepackage{graphicx}
\usepackage{amssymb}
\usepackage{enumerate}
\usepackage{multirow}
\usepackage{listings}
\usepackage{url}
\usepackage{rotating}
\usepackage{wasysym}
\usepackage{footmisc}

\newtheorem{example}{Example}

\newtheorem{definition}{Definition}

\def\etal{\textit{et al.}}

\def\toolname{Y2U}
\def\yakindu{Yakindu}
\def\uppaal{UPPAAL}

\def\toolweb{\url{www.cs.iit.edu/~code/software/Y2U/index.html}}

\def\medicalGuideline{\mathcal{G}}
\def\map{\mathcal{M}}
\def\medicalAction{a}
\def\medicalActionSet{A}
\def\resSet{R}
\def\res{r}
\def\state{S}
\def\transition{T}
\newcommand{\resVar}[1]{V_{#1}} 
\def\guard{G}

\begin{document}

\title{Model and Integrate Medical Resource Availability into Verifiably Correct Executable Medical Guidelines - Technical Report}

\author{
	\IEEEauthorblockN{Chunhui Guo, Zhicheng Fu, Zhenyu Zhang, Shangping Ren\thanks{We would like to thank Professor Yu Jiang from Tsinghua University for his valuable suggestions in this work. The work is supported in part by NSF CNS 1545008 and NSF CNS 1545002.}}
	\IEEEauthorblockA{Department of Computer Science\\
		Illinois Institute of Technology\\
		Chicago, IL 60616, USA\\
		Email: \{cguo13, zfu11, zzhang111\}@hawk.iit.edu, ren@iit.edu}	
	\and
	\IEEEauthorblockN{Lui Sha}
	\IEEEauthorblockA{Department of Computer Science\\
		University of Illinois at Urbana-Champaign\\
		Urbana, IL 61801, USA\\
		Email: lrs@illinois.edu}
}

\maketitle

\begin{abstract}
Improving effectiveness and safety of patient care is an ultimate
objective for medical cyber-physical systems.
A recent study shows that the patients' death rate can be reduced by
computerizing medical guidelines~\cite{Mckinley2011computer}.
Most existing medical guideline models are validated and/or verified
based on the assumption that all necessary medical resources needed
for a patient care are always available.
However, the reality is that some medical resources,
such as special medical equipment or medical specialists,
can be temporarily unavailable for an individual patient.
In such cases, safety properties
validated and/or verified in existing medical guideline models without considering
medical resource availability may not hold any more.

The paper argues that considering medical resource availability is essential
in building verifiably correct executable medical guidelines.
We present an approach to explicitly
and separately model medical resource availability and automatically integrate
resource availability models into an existing statechart-based computerized
medical guideline model.
This approach requires minimal change in existing medical guideline models
to take into consideration of medical resource availability
in validating and verifying medical guideline models.
A simplified stroke scenario
is used as a case study to investigate the effectiveness and validity of our approach.

\end{abstract}

\section{Introduction and Related Work}
\label{sec:intro}
Medical guidelines play an important role in today's
medical care. Over past two decades,
significant amount of efforts have also been made in obtaining various
computer-interpretable models and developing tools for the management of medical guidelines,
such as Asbru~\cite{Balser2002Asbru}, GLIF~\cite{patel1998representing},
GLARE~\cite{Terenziani2004GLARE}, EON~\cite{Tu2001EON}, and
PROforma~\cite{fox1998disseminating}.
Along with the well development and use of formal techniques on system design \cite{jiang2015design,jiang2016stateflow,yang2016verifying},
our previous work~\cite{Guo2016ICCPS} also designed a platform to model medical guidelines
with statecharts and automatically transform 
statecharts~\cite{harel1987statecharts} to timed automata~\cite{alur1994theory} for formal verification.
Furthermore, runtime verification is proposed and well adapted to working directly on the medical guidance
systems \cite{jiang2017data,jiang2016use,Guo2017COMPSAC} to improve the system performance.
All these work is based on medical guidelines presented in medical handbooks.

However, medical guidelines often focus on medical procedures
and with implicit assumption
that all required medical resources for treatments are always available.
By medical resources, we mean medical professionals, supplies, and
equipments\footnote{Patients are not considered as medical resources.
They can be
treated as preconditions of treatments and can be
validated with the protocol presented in~\cite{WuTreatment2014}.}.
Most existing computer-based medical guideline models inherit the implicit assumption
and are validated and/or verified based on that
all required medical resources are constantly available.
Unfortunately, the reality is that some medical resources, such as special medical equipments
or medical specialists, can be temporarily unavailable for patients.
In such cases, some processes of medical guideline models may be blocked and safety properties
validated and/or verified may fail and put patients in danger.
We use a simplified stroke scenario
to illustrate the cases
as follows.
For illustration purpose, we ignore some medical details from computer science
perspective in the simplified stroke scenario.

\textbf{Stroke Scenario:}\textit{
	An ischemic stroke occurs when a clot or a mass blocks a blood vessel, cutting
	off blood flow to a part of the brain and results in a corresponding loss
	of neurologic function~\cite{IschemicStroke}.
	The goal of acute treatment is to keep the amount of brain injury as
	small as possible.	
	The only FDA approved drug to treat ischemic stroke is
	tissue plasminogen activator (tPA), a clot busting drug~\cite{IschemicStroke}.	
	The intravenous (IV) tissue plasminogen activator (tPA) injection is a standard treatment for ischemic stroke patients and it is most effective during the initial 3-hour window from the onset of stroke symptoms~\cite{tpa-gold-standard}. The treatment window can be extended from 3 to 4.5 hours for certain patients, but the risks are increased~\cite{StrokeGuideline}. Some patients can be treated by dripping tPA directly on the clot through a intra-arterial (IA) micro-catheter within 6 hours from the onset of stroke symptoms~\cite{Prince2013strokeIA}. However, the IA tPA treatment requires specialists to control tPA dose, special equipments to put the micro-catheter into blood vessels, and technicians to operate the special equipment.}

\textit{In addition, in order to use the tPA treatment, we must ensure that (1) CT scan does not show hemorrhage,  and (2) the patient's blood pressure is under control. To derive the conclusion that the patient does not have hemorrhage, we would need medical resources including a CT machine, a CT technician, and a radiologist. If a patient's blood pressure is not within the range for tPA administration, a specialist is required to control blood pressure.} 

In the simplified stroke scenario, there are three medical properties needed to be guaranteed in the patient care:
\begin{itemize}
	\item \textbf{P1}: the tPA is injected only if a CT scan shows no hemorrhage
	and systolic and diastolic blood pressure are smaller than or equal to
	185 mm Hg and 110 mm Hg;
	
	\item \textbf{P2}: the IV tPA administration is completed within
	3 hours from onset of symptoms;
	
	\item \textbf{P3}: the IA tPA administration is
	completed within 6 hours from onset of symptoms.
\end{itemize}
Assume a stroke patient's onset time is 0 and a physician
orders CT scan for the patient at time 20 (minutes). If the CT machine is always available,
the tPA administration can be completed within the 3-hour window.
However, if the CT machine is unavailable until 200 minutes.
In such case, the tPA administration can not be completed within 3 hours
due to temporarily unavailable CT machines.
Hence, modeling medical resources in existing medical guideline models and
validating and verifying safety properties with consideration of medical resource availability
are essential for improving patient care safety.

One approach to address the medical resource availability issue
in existing medical guideline models is to directly add
medical resource availability as guards
to corresponding transitions or as state constraints.
We call this method as direct modification approach.
The timed and resource-oriented statecharts~\cite{Kim2010TII} takes
the direct modification approach by specifying required resource information in states.
Christov \etal~\cite{Christov2008Formally,Christov2008} uses Little-JIL to model the processes in medical guidelines and represents resource as preconditions of process steps.
The mentioned work has shown that adding medical resource availability as
transition guards, state constraints, or process preconditions
is a practicable approach to address medical resource temporal unavailable issue in medical guideline models.
But these approaches also face the following challenges.
First often times, a medical guideline represents a generalized treatment procedure for a disease, it is not defined for a specific hospital. As medical resource availabilities at different medical facilities can be significantly different, to use such direct modification approaches, we would have to build different medical guideline models for
different medical facilities.
Second, even within a same medical facility, medical resource availability
can change over time, therefore the corresponding medical guideline models
need to be changed as well.
Third, for a failed safety property, identifying the errors that cause the failure becomes more challenging as errors both in medical resource availabilities and medical guideline model itself could cause the safety property to fail.
Forth, medical guideline models with medical resource built in would increase the difficulty for medical professionals to understand and clinically validate the models, and would unnecessarily require medical staffs to know the medical resource availability at medical facilities.

In this paper, we present an approach to model and integrate medical resource availability into executable medical guideline models. Our approach separates resource models from medical guideline models to minimize the change impact of both guidelines and resources, as well as leaving the syntax and semantics of medical guideline models unchanged. In particular, we first define the procedures that how physicians to annotate required resources for actions in medical guideline models. To explicitly take medical resource availability into medical guideline system design, we represent an approach to explicitly and separately model medical resource availability. The medical resource availability models are then integrated into medical guideline models so that the integrated medical guideline models can be validated  and safety properties in the presence of temporarily unavailable resources can be formally verified.
A simplified stroke scenario
is used as a case study to explain the proposed approach. The main contributions of the paper are:
\begin{itemize}
	\item Take medical resource availability into consideration in validating and verifying executable medical guideline models.	
	\item Present an approach to explicitly and separately model medical resource availability with statecharts.
	\item Develop an approach to automatically integrate resource availability models with verifiably correct executable medical guideline models.
\end{itemize}

The rest of the paper is organized as following: we introduce a framework for building verifiably correct executable medical guideline models in Section~\ref{sec:statechart}. Section~\ref{sec:resource} describes the approaches for explicitly and separately modeling medical recourses and their variabilities. Section~\ref{sec:integration} defines the procedure for integrating medical resource availability models into medical guideline models.
A simplified stroke case study is given in Section~\ref{sec:exp} to 
illustrate the effectiveness of the presented approach.
We draw conclusions and point out future work in Section~\ref{sec:conclusion}.

\section{Verifiably Correct Executable Medical Best Practice Guidelines}
\label{sec:statechart}
\begin{figure}[ht]
	\centering
	\includegraphics[width = 0.4\textwidth]{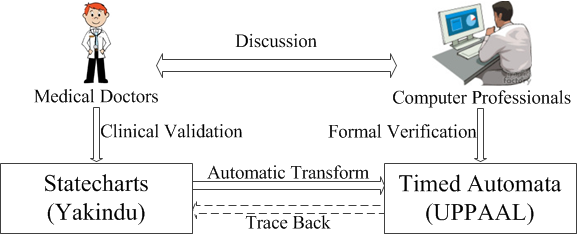}
	\caption{A Platform for Building Verifiably Correct Executable Medical Guidelines}
	\label{fig:approach}
\end{figure}

Our previous work~\cite{Guo2016ICCPS} designed a platform to build verifiably
correct executable medical guidelines.
The high level abstract of the platform is depicted in Fig.~\ref{fig:approach}.
In particular,
we use statecharts~\cite{harel1987statecharts} to model medical guidelines and
interact with medical professionals to validate the correctness of the medical
guideline models.
The statecharts built with \yakindu\ tool~\cite{yakindu}
are then automatically transformed to
timed automata~\cite{alur1994theory} by the developed \toolname
\footnote{\label{footnote:y2u}The \toolname\ tool is available at \toolweb.} tool,
so that the safety properties required by the model,
\uppaal\ timed automata~\cite{behrmann2004tutorial} in particular,
can be formally verified.
We use the simplified stroke scenario presented in Section~\ref{sec:intro}
as an example to illustrate our previous approaches on how to build verifiably
correct executable medical guidelines.

\begin{figure}[ht]
	\centering
	\includegraphics[width = 0.49\textwidth]{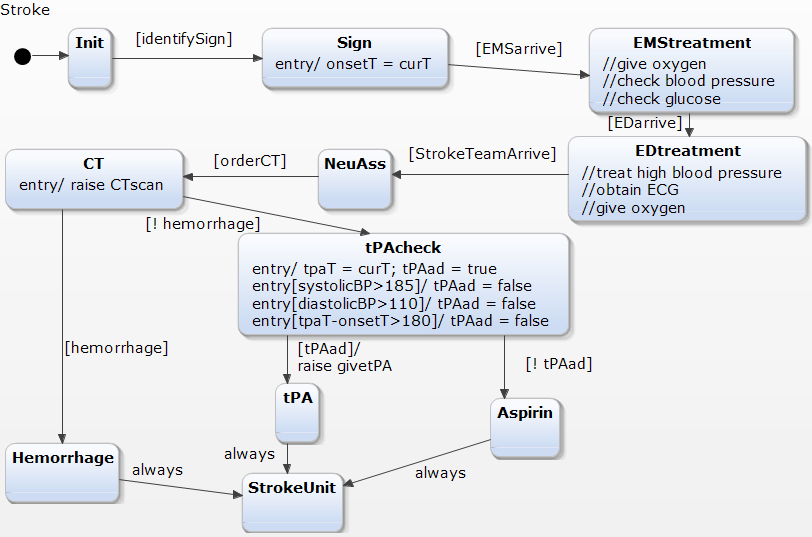}
	\caption{Simplified Stroke \yakindu\ Statechart Model}
	\label{fig:stroke}
\end{figure}

We use \yakindu\ statecharts to model the stroke treatment guideline~\cite{StrokeGuideline}.
For illustration and easy understanding purpose, we show a simplified stroke
statechart model in Fig.~\ref{fig:stroke}, which only focuses on the CT scan
and IV tPA administration procedures and omits details of other medical procedures.
The full version of stroke statechart model is available in the case study (Section~\ref{sec:exp}).
Hypertension is present in up to 84\% of patients presenting with acute
stroke~\cite{McManus2016JCN}. In the simplified statechart shown in
Fig.~\ref{fig:stroke}, we assume that upon patient arrival, treatments
to control blood pressure have been immediately performed.
A patient’s blood pressure is either quickly brought within the range or
not possible.

In the statechart shown in Fig.~\ref{fig:stroke}, two medical actions $\mathtt{CTscan}$
and $\mathtt{givetPA}$ are modeled by \yakindu\ statechart \textit{events}.
In \yakindu\ statecharts, \textit{events} can be raised
by both states and transitions.
For instance, the \textit{entry action} of state ``CT'' ($\mathtt{entry/ \ raise \ CTscan}$)
raises \textit{event} $\mathtt{CTscan}$ when state ``CT'' is entered.
The \textit{event} $\mathtt{givetPA}$ is raised by the transition from state
``tPAcheck'' to state ``tPA'' if tPA is administrated (the value of
boolean variable $\mathtt{tPAad}$ is $\mathtt{true}$).
In the simplified stroke statechart model (Fig.~\ref{fig:stroke}),
the two two timing related variables $\mathtt{curT}$ and $\mathtt{onsetT}$
represent the current system time and the onset time of stroke symptoms, respectively.
We assume that the time unit in the simplified stroke statechart model is minute.
Hence, the remaining time of the 3-hour tPA treatment window can be calculated
by formula $180 - (\mathtt{curT} - \mathtt{onsetT})$.

\begin{figure}[ht]
	\centering
	\includegraphics[width = 0.4\textwidth]{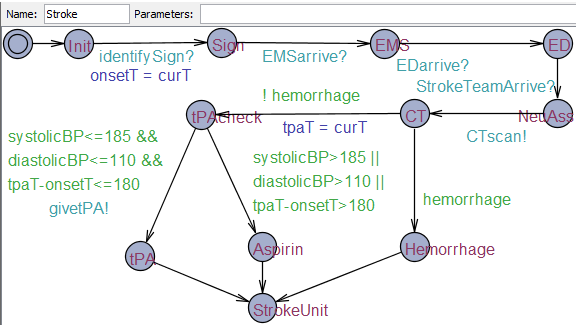}
	\caption{Simplified Stroke \uppaal\ Model}
	\label{fig:strokeU}
\end{figure}

The simplified stroke statechart model in Fig.~\ref{fig:stroke}
is transformed to \uppaal\ time automata as shown in Fig.~\ref{fig:strokeU}
with our \toolname\footref{footnote:y2u} tool~\cite{Guo2016ICCPS}.
The properties \textbf{P1} and \textbf{P2} are verified in \uppaal\ by
formula~\eqref{eq:P1} and formula~\eqref{eq:P2}, respectively.

\begin{align}
\label{eq:P1}
\begin{split}
A[~] \ \mathtt{Stroke.tPA} \ imply \ \mathtt{systolicBP}<=185 \ \&\& \\
 \mathtt{diastolicBP}<=110 \ \&\& \ ! \ \mathtt{hemorrhage}		
\end{split}
\end{align}

\begin{align}
\label{eq:P2}
A[~] \ \mathtt{Stroke.tPAcheck} \ imply \ \mathtt{tpaT}-\mathtt{onsetT}<=180
\end{align}

As medical guideline handbooks often assume that all required medical resources
are available. With the assumption, both clinical validation
results of stroke \yakindu\ model in Fig.~\ref{fig:stroke} and
formal verification results of stroke \uppaal\ model in Fig.~\ref{fig:strokeU}
show that both properties \textbf{P1} and \textbf{P2} are satisfied.
However, the assumption on medical resource availability may not always hold in reality.
For example, the $\mathtt{CTscan}$ medical action in state ``CT'' of the
stroke statechart model in Fig.~\ref{fig:stroke} requires CT machines and CT
technicians. If both CT machines and CT technicians are available after
200 minutes from onset of the symptoms, the stroke statechart is then blocked
at state ``CT'' for 200 minutes. In this scenario, the safety property
\textbf{P2} fails.

The example reveals a fact that
safety properties validated and verified in medical guideline models based on the assumption
that medical resources are always available can fail because of temporarily unavailable resources.
Hence, taking into consideration of
medical resource availability in developing verifiable medical guideline models
is essential in validating
and verifying the safety properties of the guideline models.
We model medical resource availability with statecharts
and integrate medical resource availability models
with medical guideline statecharts to validate and verify safety properties
in the following two sections. 

\section{Model Medical Resource Availability with Statecharts}
\label{sec:resource}
In this section, we model medical resource availability with statecharts in
two steps:
(1) automatically annotate required medical resources
in executable medical guidelines
and
(2) explicitly model medical resource availability with statecharts based
on resource annotations and given availability information.

\subsection{Annotate Medical Resources in Executable Medical Guideline Models}
\label{subsec:annotation}

To model medical resource availability, we need to identify which resources
are required by which medical actions and represent
the required resources in executable medical guidelines.

We use the simplified stroke statechart model shown in Fig.~\ref{fig:stroke} as
an example to illustrate medical resources required by medical guidelines.
In the state ``CT'', a medical action $\mathtt{CTscan}$ which is modeled
as an \textit{event} in \yakindu\ statecharts is raised by the \textit{entry
action} of the ``CT'' state. According to medical professionals, the $\mathtt{CTscan}$
medical action requires CT machines and CT technicians.
Similarly, a medical action $\mathtt{givetPA}$ is raised by the \textit{action}
of the \textit{transition} from state ``tPAcheck'' to state ``tPA''.
The $\mathtt{givetPA}$ medical actioin requires tPA fluid.
The examples show that
(1) medical actions are modeled as statechart \textit{events} and can be raised
in both \textit{states} and \textit{transitions}
and
(2) medical resources required by medical actions are implicit and not represented
in medical guideline statecharts.
They need to be provided by medical professionals.

As medical professionals participate in model building and clinical
validation processes of medical guideline statecharts, one intuitive
method to represent required resources in medical guideline statecharts
is that medical professionals review each \textit{state} and \textit{transition}
of medical guideline statecharts and manually annotate required medical
resources in each \textit{state} and \textit{transition}.
The intuitive method works but has a disadvantages that
medical professionals need to check all \textit{states} and \textit{transitions}
in guideline statecharts when validating the correctness of annotated
resource information.

To avoid the disadvantage, we propose an approach to annotate medical
resources in executable medical guidelines with two steps:
(1) represent medical actions required resources given by medical professionals
by a map structure
and
(2) automatically annotate required medical resources in \textit{states}
and \textit{transitions} according to the resource map and raised medical
actions in corresponding \textit{states} and \textit{transitions}.
Compared to the above intuitive resource annotation method, the proposed approach
has an advantage that medical professionals only need to check the resource map
when validating the correctness of medical resource information.

In the resource map structure $(\mathtt{key}, \overrightarrow{\mathtt{value}})$,
the \textit{key} is medical actions that are
represented by corresponding \textit{event} names in the medical guideline statecharts.
The \textit{value} of the resource map is required medical resources of the
corresponding \textit{key} (medical action). As a medical action may require
multiple resources, we use an array of all required medical resources to
represent the \textit{value} in the resource map structure.
In the resource array, we replace spaces in resource names with underscores ($\_$).
In current work, we only consider the multiple resources required by the same medical
action are pairwise independent and leave dependent resources as our future work.
As multiple resources are independent, the sequence of multiple resources
in a resource array is not important.
We give the formal definition of the resource map structure in Definition~\ref{def:map}
and show the resource map of the simplified stroke scenario in Example~\ref{ex:map}.

\begin{definition}
	\label{def:map}
	Given an executable medical guideline model $\medicalGuideline$, a set
	of medical actions $\medicalActionSet = \{ \medicalAction_1,
	\medicalAction_2, \dots, \medicalAction_n \}$ in the medical guideline $\medicalGuideline$,
	and a set of medical resources $\resSet_i = \{ \res_1^i, \res_2^i, \dots, \res_m^i\}$
	required by the medical action $\medicalAction_i$,
	the medical resource map $\map$ is defined as
	\begin{align}
	\label{eq:map}
		\begin{split}		
		\map = &\{ (\medicalAction_1, [\res_1^1, \res_2^1, \dots, \res_{m_1}^1]), \\		
		&(\medicalAction_2, [\res_1^2, \res_2^2, \dots, \res_{m_2}^2]), \\		
		&\dots \dots \dots \\
		&(\medicalAction_n, [\res_1^n, \res_2^n, \dots, \res_{m_n}^n]) \}.
		\end{split}
	\end{align}
\end{definition}

\begin{example}
	\label{ex:map}
	The simplified stroke statechart model shown in Fig.~\ref{fig:stroke}
	has two medical actions $\mathtt{CTscan}$ and $\mathtt{givetPA}$.
	Suppose the $\mathtt{CTscan}$ medical action requires CT machines
	and CT technicians and the $\mathtt{givetPA}$ medical actioin requires
	tPA. According to Definition~\ref{def:map}, the resource map of
	the simplified stroke scenario is
	\begin{align}
	\label{eq:mapStroke}
	\{ (\mathtt{CTscan}, [\mathtt{CT\_machine}, \mathtt{CT\_technician}]),
	(\mathtt{givetPA}, [\mathtt{tPA}]) \}.
	\end{align}	
\end{example}

The required medical resource information represented in the map $\map$ is
independent of executable medical guideline models.
To model medical resource availability, we also need to annotate
the required resources in executable medical guideline models.
With the purpose of not affecting execution behaviors and validation/verification
results of medical verifiably correct executable medical guideline models, we annotate
medical resources by \yakindu\ statechart \textit{comments}. The annotation
is defined as follows.
\begin{definition}
	\label{def:annotation}
	Given a state $\state$ (or a transition $\transition$) in a executable medical
	guideline model $\medicalGuideline$,
	a set of medical actions $\medicalActionSet_S = \{ \medicalAction_1,
	\medicalAction_2, \dots, \medicalAction_k \}$ modeled in state $\state$
	(or transition $\transition$), and	
	a medical resource map $\map$ of $\medicalGuideline$,
	the annotation of state $\state$ (or transition $\transition$) is represented as
	\begin{align}
	\label{eq:annotation}
	//@\mathtt{RES}: \res_1^1, \dots, \res_{m_1}^1, \dots, \res_1^2, \dots, \res_{m_2}^2
	, \dots, \res_1^k, \dots, \res_{m_k}^k.
	\end{align}	
\end{definition}

Based on the medical resource map and the
medical resource annotation definitions, we
annotate required medical resources in executable medical guideline statecharts with
following two steps:
first search each \textit{state} $\state$ (and transition $\transition$) in the given
medical guideline statechart $\medicalGuideline$;
second, if the actions of state $\state$ (or transition $\transition$) contain medical actions
in the given medical resource map $\map$, add annotation, i.e.,
formula~\eqref{eq:annotation}, to state $\state$ (or transition $\transition$).
Algorithm~\ref{alg:annotation} gives the details of the annotation procedure,
where the operation $\resSet + \resSet'$ in Line 6
returns the concatenation of $\resSet$ and $\resSet'$.
The time complexity of Algorithm~\ref{alg:integration} is $O(L*M*N)$, where
$L$ is the element number of the medical resource map $\map$,
$M$ is the number of medical resources required by the medical guideline model $\medicalGuideline$,
and $N$ is the sum of states' number and transitions' number in $\medicalGuideline$.

\begin{algorithm}
	\caption{\textsc{Annotation}}
	\label{alg:annotation}
	\begin{algorithmic}[1]
		\REQUIRE An executable medical guideline model $\medicalGuideline$ and
		 a medical resource map $\map$ (formula~\eqref{eq:map}).
		\ENSURE The annotated medical guideline model $\medicalGuideline'$.
		
		\FOR{each state $\state$ or transition $\transition$ in $\medicalGuideline$}
			\STATE Define a resource array $\resSet = [~]$
			\FOR{each raised action $\medicalAction$ in $\state$ or $\transition$}
				\STATE Find $\resSet'$ with key $\medicalAction$ in $\map$
				\IF{$\resSet'$ is not $\mathtt{NULL}$}
					\STATE $\resSet = \resSet + \resSet'$
				\ENDIF			
			\ENDFOR
			\IF{$\resSet$ is not empty}						
				\STATE Add an annotation in the format of formula~\eqref{eq:annotation} to state $\state$ or transition $\transition$
			\ENDIF
		\ENDFOR
		\RETURN $\medicalGuideline$
	\end{algorithmic}
\end{algorithm}

\begin{example}
	\label{ex:strokeAnnotation}
	Given the simplified stroke statechart model shown in Fig.~\ref{fig:stroke}
	and a resource map of formula~\eqref{eq:mapStroke}.	
	The state ``CT'' has a medical action $\mathtt{CTscan}$.
	We use $\mathtt{CTscan}$ as the key to search the resource map given by
	formula~\eqref{eq:mapStroke} and find resource array
	$[\mathtt{CT\_machine}, \mathtt{CT\_technician}]$. According to
	Definition~\ref{def:annotation}, we add the annotation
	``$//@\mathtt{RES}: \mathtt{CT\_machine}, \mathtt{CT\_technician}$''
	to state ``CT''.
	Similarly, we add the annotation ``$//@\mathtt{RES}: \mathtt{tPA}$''
	to the transition from state ``tPAcheck'' to state ``tPA''.
	The annotated stroke statechart	model by Algorithm~\ref{alg:annotation}
	is depicted in Fig.~\ref{fig:strokeAnnotation}, where the annotated states and
	transitions are marked by red rectangle.

	\begin{figure}[ht]
		\centering
		\includegraphics[width = 0.49\textwidth]{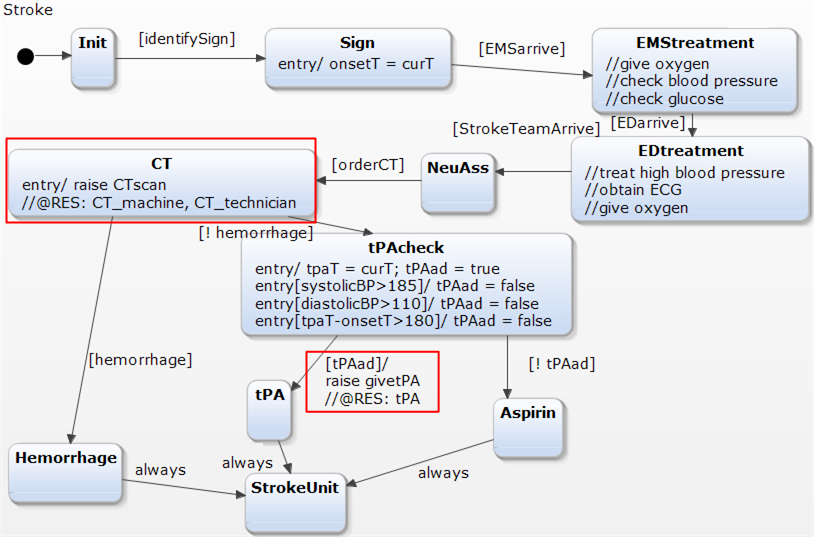}
		\caption{Annotated Stroke \yakindu\ Model}
		\label{fig:strokeAnnotation}
	\end{figure}
\end{example}

\subsection{Model Medical Resource Availability with Statecharts}
\label{subsec:resource}

Given a resource map $\map$ and resource availability information,
we develop statecharts to model medical resource availability
in three steps:
(1) design a $\mathtt{Timer}$ statechart to record current system time;
(2) declare a boolean variable for each resource to denote its availability at
current time;
and
(3) build a statechart for each resource to represent its given availability information.

\begin{figure}[ht]
	\centering
	\includegraphics[width = 0.2\textwidth]{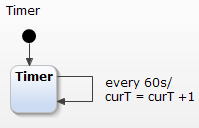}
	\caption{Timer Statechart}
	\label{fig:timer}
\end{figure}

For the $\mathtt{Timer}$ statechart,
we use an integer variable $\mathtt{curT}$ to denote current system time
and let a $\mathtt{Timer}$ statechart to increase current time $\mathtt{curT}$.
The $\mathtt{Timer}$ statechart only contains one state which has a self-loop
transition to increase current time $\mathtt{curT}$ by 1 every one time unit.
Fig.~\ref{fig:timer} shows an example $\mathtt{Timer}$ statechart with time
unit minute, which increases $\mathtt{curT}$ by 1 $\mathtt{every \ 60s}$.

\begin{figure}[ht]
	\centering
	\includegraphics[width = 0.3\textwidth]{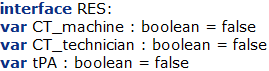}
	\caption{Resource Availability Variables}
	\label{fig:interface}
\end{figure}

To represent resource availability related variables, we declare an interface
named $\mathtt{RES}$. For each unique resource $\res$ in a given resource map $\map$,
we declare a boolean variable $\resVar{\res}$ in the interface $\mathtt{RES}$ to
denote the resource $\res$'s availability at current system time.
The variable $\resVar{\res}$ has the same name as the corresponding resource $\res$
and default value $\mathtt{false}$ that means the resource $\res$ is not
available initially.
For example, the resource map of the simplified stroke
scenario given in formula~\eqref{eq:mapStroke} contains three medical resources
$\mathtt{CT\_machine}$, $\mathtt{CT\_technician}$, and $\mathtt{tPA}$.
The declared resource availability variables of the simplified stroke
scenario is shown in Fig.~\ref{fig:interface}.

For each unique resource $\res$ in the given resource map $\map$, we build a
statechart to represent its given availability information.
Each resource statechart contains only one state $\state$ that has a self-loop
transition $\transition$ with guard $\mathtt{true}$. The transition $\transition$ ensures
that the resource $\res$'s availability is checked at each statechart
execution cycle and maintains the latest value.
The entry actions of the state $\state$ check the resource $\res$'s
availability at current time $\mathtt{curT}$ based on given resource
availability information.
If the resource $\res$ is available, the entry action assigns $\mathtt{true}$
value to the corresponding resource boolean variable $\resVar{\res}$;
otherwise, the resource variable $\resVar{\res}$ is assigned as $\mathtt{false}$.
We use Example~\ref{ex:resource} to show resource statecharts for the
simplified stroke scenario.

\begin{example}
	\label{ex:resource}
	For the simplified stroke scenario, given the resource map as
	formula~\eqref{eq:mapStroke} and resource availability information as
	follows:
	(1) both $\mathtt{CT\_machine}$ and $\mathtt{CT\_technician}$ are available
	after 200 minutes
	and
	(2) the $\mathtt{tPA}$ is always available.	
	For resource $\mathtt{CT\_machine}$,
	we build the ``CT\_machine'' statechart with only one state named
	``CT\_machine'' which has a self-loop transition with guard $\mathtt{true}$.
	According to given resource availability information that
	the $\mathtt{CT\_machine}$ is available after 200 minutes, we add two
	entry actions to the state ``CT\_machine'':
	\begin{enumerate}
		\item $\mathtt{entry[curT>200]/ RES.CT\_machine = true}$ assigns
		variable $\mathtt{CT\_machine}$ as $\mathtt{true}$ if current time
		$\mathtt{curT}$ is larger than 200 minutes and denotes that
		the resource $\mathtt{CT\_machine}$ is available after 200 minutes;
		
		\item $\mathtt{entry[curT<=200]/ RES.CT\_machine = false}$ assigns
		variable $\mathtt{CT\_machine}$ as $\mathtt{false}$ if current time
		$\mathtt{curT}$ is smaller than or equal to 200 minutes and denotes that
		the resource $\mathtt{CT\_machine}$ is not available until 200 minutes.
	\end{enumerate}
	Similarly, we build two statecharts for resource $\mathtt{CT\_technician}$ and $\mathtt{tPA}$, respectively. The resource statecharts are shown in Fig.~\ref{fig:resoruce}.	

	\begin{figure}[ht]
		\centering
		\includegraphics[width = 0.3\textwidth]{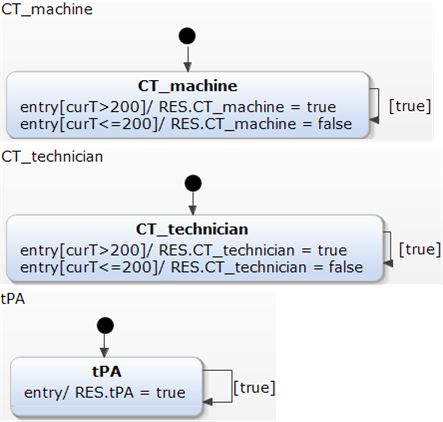}
		\caption{Resource Statecharts}
		\label{fig:resoruce}
	\end{figure}
	
\end{example}

\section{Integrate Medical Resource Availability Models with Medical Guideline Statecharts}
\label{sec:integration}
To clinically validate and formally verify the safety of medical guideline models
with consideration of medical resource availability, we need to integrate medical
resource availability models with medical guideline statecharts.

According to the medical resource availability modeling approach presented
in Section~\ref{subsec:resource}, for each resource $\res$,
a boolean variable $\resVar{\res}$ is declared.
We use the declared resource availability variable $\resVar{\res}$ to
bridge the communication between
medical resource availability models and medical guideline statecharts
and modify medical guideline statecharts with following integration rules.
\begin{itemize}
	\item \textbf{Integration Rule 1}: For each transition $\transition$ with
	guard $\guard$, if it is annotated by
	``$//@\mathtt{RES}: \res_1, \res_2, \dots, \res_n$'',
	the guard $\guard$ is modified by	
	$G = G \ \&\& \ \resVar{\res_1} \ \&\& \ \resVar{\res_2} \ \&\& \ \dots \ \&\& \ \resVar{\res_n}$;
	
	\item \textbf{Integration Rule 2}: For each state $\state$, if it is annotated
	with ``$//@\mathtt{RES}: \res_1, \res_2, \dots, \res_n$'',
	apply \textbf{Integration Rule 1} to all incoming transitions of the state $\state$
	with the annotation.
\end{itemize}
Algorithm~\ref{alg:integration} gives the integration procedure.
The time complexity of Algorithm~\ref{alg:integration} is $O(M*N^2)$, where
$M$ is the number of medical resources required by the medical guideline model $\medicalGuideline$
and $N$ is the sum of states' number and transitions' number in $\medicalGuideline$.
Example~\ref{ex:integration} illustrates how we apply the integration rules to
integrate the CT machine, CT technician, and tPA fluid availability models with the
simplified stroke statechart.

\begin{algorithm}
	\caption{\textsc{Integration}}
	\label{alg:integration}
	\begin{algorithmic}[1]
		\REQUIRE An annotated medical guideline model $\medicalGuideline$.
		\ENSURE The integrated medical guideline model $\medicalGuideline'$.		
		
		\FOR{each state $\state$ in $\medicalGuideline$}
			\IF{$\state$ is annotated with ``$//@\mathtt{RES}: \res_1, \res_2, \dots, \res_n$''}
				\FOR{each incoming transition $\transition$ with guard $\guard$ of state $\state$}
					\STATE $G = G \ \&\& \ \resVar{\res_1} \ \&\& \ \resVar{\res_2} \ \&\& \ \dots \ \&\& \ \resVar{\res_n}$
				\ENDFOR
			\ENDIF
		\ENDFOR
				
		\FOR{each transition $\transition$ with guard $\guard$ in $\medicalGuideline$}
			\IF{$\transition$ is annotated with ``$//@\mathtt{RES}: \res_1, \res_2, \dots, \res_n$''}	
				\STATE $G = G \ \&\& \ \resVar{\res_1} \ \&\& \ \resVar{\res_2} \ \&\& \ \dots \ \&\& \ \resVar{\res_n}$	
			\ENDIF
		\ENDFOR		
		\RETURN $\medicalGuideline$
	\end{algorithmic}
\end{algorithm}

\begin{example}
\label{ex:integration}
	We integrate the resource availability models in Fig.~\ref{fig:resoruce}
	with the annotated stroke statechart model in Fig.~\ref{fig:strokeAnnotation}.		
	The transition $\transition_1$ from state ``tPAcheck''
	to state ``tPA'' is annotated with ``$\mathtt{//@RES: tPA}$''
	and has guard $\guard_1 = \mathtt{tPAad}$. Based on \textbf{Integration Rule 1},
	the transition $\transition_1$' guard is set as
	$\guard_1 = \mathtt{tPAad} \ \&\& \ \mathtt{RES.tP}A$.
	The state ``CT'' is annotated by ``$\mathtt{//@RES: CT\_machine, CT\_technician}$'' 
	and only has one incoming transition $\transition_2$ with guard
	$\guard_2 = \mathtt{orderCT}$ from state ``NeuAss''.
	According to \textbf{Integration Rule 2}, we apply
	\textbf{Integration Rule 1} to the transition $\transition_2$ and
	set the guard as $\guard_2 = \mathtt{orderCT} \ \&\& \ \mathtt{RES.CT\_machine}  \ \&\& \ \mathtt{RES.CT\_technician}$.
	Fig.~\ref{fig:strokeIntegration} shows the integrated stroke statechart,
	where the modified transitions are marked by red rectangle.

	\begin{figure}[ht]
		\centering
		\includegraphics[width = 0.49\textwidth]{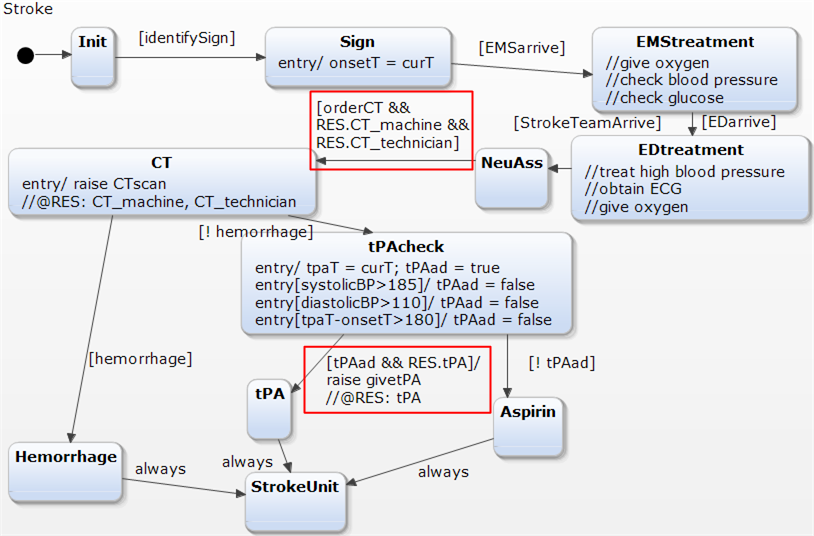}
		\caption{Integrated Stroke \yakindu\ Model}
		\label{fig:strokeIntegration}
	\end{figure}	
\end{example}

\begin{figure}[ht]
	\centering
	\includegraphics[width = 0.49\textwidth]{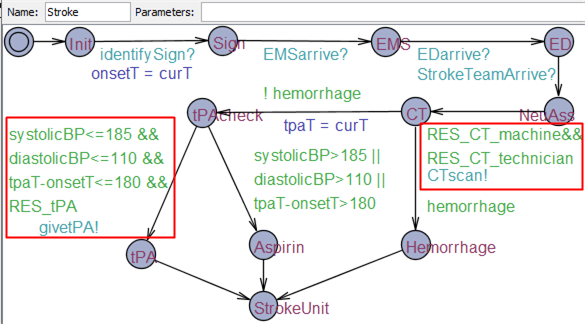}
	\caption{Integrated Stroke \uppaal\ Model}
	\label{fig:strokeIntegrationU}
\end{figure}

To clinically validate and formally verify the safety of the stroke
statechart with the consideration of resource availability,
we run simulations of the integrated stroke model (Fig.~\ref{fig:strokeIntegration})
through \yakindu, transform it to integrated stroke \uppaal\ model
(Fig.~\ref{fig:strokeIntegrationU}), and verify the two safety properties
(\textbf{P1} and \textbf{P2}) in \uppaal.
The resource availability is given in Example~\ref{ex:resource}, i.e.,
both $\mathtt{CT\_machine}$ and $\mathtt{CT\_technician}$ are available
after 200 minutes, the $\mathtt{tPA}$ is always available.
Both simulation and verification results show that
the property \textbf{P1} holds while \textbf{P2} fails.

\section{Simplified Stroke Case Study}
\label{sec:exp}
The stroke statechart model given in Fig.~\ref{fig:stroke} has only focused on the CT scan and IV tPA administration procedures, but omitted the details of other medical procedures. To validate the effectiveness of the proposed approach, we extend the simplified stroke model by considering following scenarios with different patient conditions: (1) if a patient's blood pressure is not within the range required by tPA administration, a blood pressure control procedure needs to be performed; (2) if tPA administration is approved within 3 hours from onset of stroke symptoms, an IV tPA procedure is performed; (3) if tPA administration is approved in the 3-6 hour window from the onset time, an IA tPA procedure is performed; and (4) if tPA is not approved, aspirin is given to patients.

\begin{figure}[ht]
	\centering
	\includegraphics[width = 0.49\textwidth]{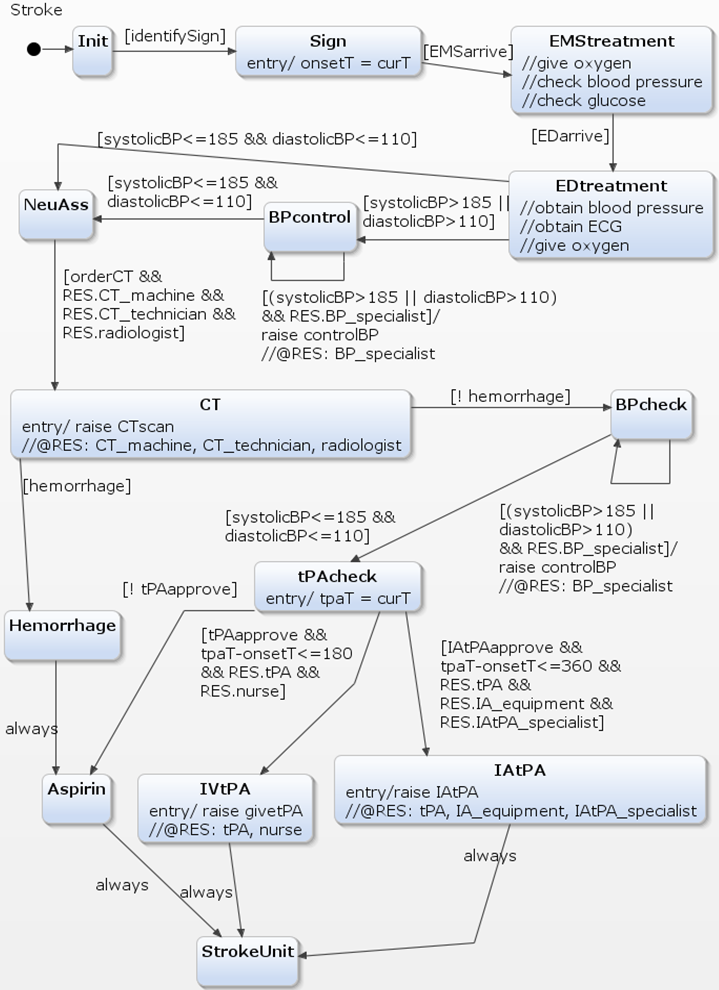}
	\caption{Integrated Stroke Statechart}
	\label{fig:strokeFull}
\end{figure}

We use the proposed approach to annotate resources, model resource
availabilities, and integrate resource models with
the extended stroke statechart model.
The integrated stroke statechart
is shown in Fig.~\ref{fig:strokeFull}.

\begin{figure}[ht]
	\centering
	\includegraphics[width = 0.49\textwidth]{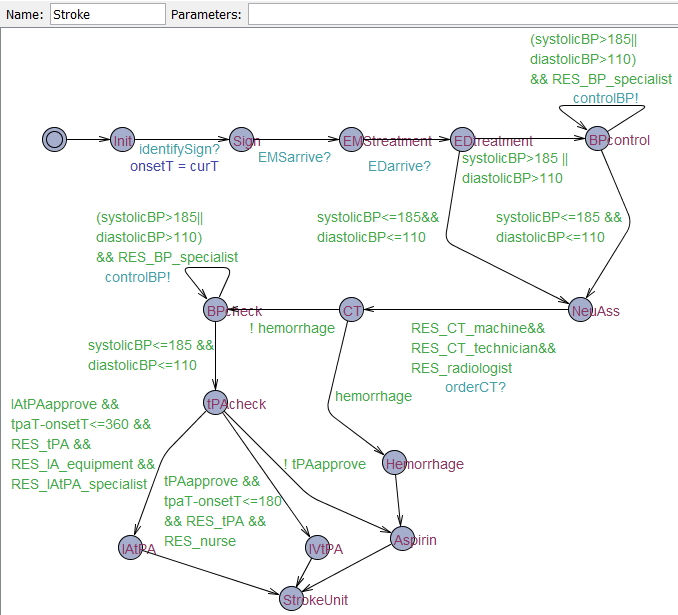}
	\caption{Stroke \uppaal\ Model}
	\label{fig:strokeFullU}
\end{figure}

To clinically validate and formally verify the safety of the stroke
statechart with the consideration of resource availabilities,
we run simulations of the integrated stroke statechart model (Fig.~\ref{fig:strokeFull})
through \yakindu, transform the integrated stroke model to an \uppaal\ model with
the \toolname\ tool~\cite{Guo2016ICCPS}
(Fig.~\ref{fig:strokeFullU}), and verify the safety properties in \uppaal.

In addition to the properties \textbf{P1} and \textbf{P2} given in
formula~\eqref{eq:P1} and formula~\eqref{eq:P2},
we also need to verify property \textbf{P3} that the IA tPA administration must
be completed within 6 hours from onset of stroke symptoms, i.e.,
\begin{align}
\label{eq:P3}
A[~] \ \mathtt{Stroke.IAtPA} \ imply \ \mathtt{tpaT}-\mathtt{onsetT}<=360.
\end{align}

\begin{figure}[ht]
	\centering
	\includegraphics[width = 0.45\textwidth]{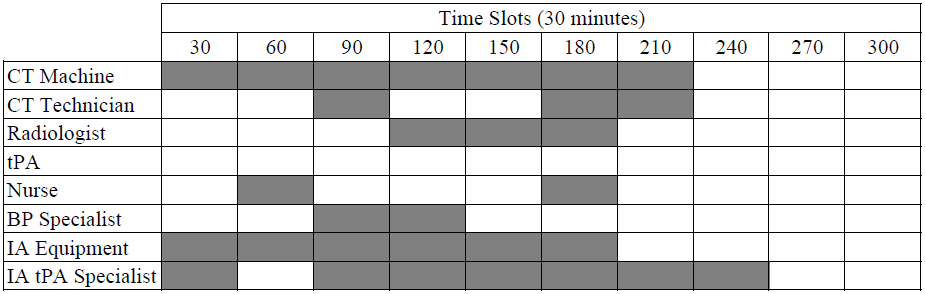}
	\caption{Stroke Resource Schedule}
	\label{fig:StrokeSchedule}
\end{figure}

Assume a patient's onset time of stroke symptom is 0,
the resource schedule is given in Fig.~\ref{fig:StrokeSchedule},
where resources are not available during shaded time slots.
Both simulation and verification results show that
the safety property \textbf{P1} and \textbf{P3} hold, but \textbf{P2} fails.

The case study demonstrates that
the proposed approach is effective in capturing
safety property fails caused by temporarily unavailable resources
in both clinical validation and formal verification process.

\section{Conclusion}
\label{sec:conclusion}
Medical guidelines often assume that all required medical resources
are available. Unfortunately, the reality at medical facilities
is that some medical resources can be temporarily unavailable.
Hence, taking into consideration of
medical resource availability in developing verifiable medical guideline models
is essential in validating and verifying safety properties.
The paper presents an approach to separately model medical resource availability
with statecharts and automatically integrate medical resource availability
statecharts with verifiably correct executable medical guideline models.
The proposed approach allows to minimize the change impact on
medical guideline models caused by
resource availability variations.
Applying separation of concern methodology in our approach further
allows different professionals to focus on only their own domains,
e.g., medical professionals and resource administrators
focus on medical guidelines and medical resource availability information,
respectively.
The separation also improves model understandability for
both medical professionals and resource administrators.
In addition, this approach can be easily implemented in
our existing platform~\cite{Guo2016ICCPS} with which
the medical guideline models with the consideration of resource availability
can be clinically validated by medical professionals and formally verified
with existing tools.
We also use a simplified stroke scenario
as a case study to investigate 
the effectiveness and validity of our approach.
In this paper, we consider the multiple resources required by the same medical
action are pairwise independent. Our future work is to extend the presented approach
to support dependent resources.

\bibliographystyle{abbrv}
\bibliography{ref}

\begin{thebibliography}{10}

\bibitem{alur1994theory}
R.~Alur and D.~L. Dill.
\newblock A theory of timed automata.
\newblock {\em Theoretical computer science}, 126(2):183--235, 1994.

\bibitem{IschemicStroke}
A.~S. Association.
\newblock Ischemic stroke.
\newblock
  \url{https://www.strokeassociation.org/idc/groups/stroke-public/@wcm/@hcm/documents/downloadable/ucm_309725.pdf}.

\bibitem{tpa-gold-standard}
A.~S. Association.
\newblock Stroke treatment.
\newblock
  \url{http://www.strokeassociation.org/STROKEORG/AboutStroke/Treatment/Stroke-Treatment_UCM_492017_SubHomePage.jsp}.

\bibitem{Balser2002Asbru}
M.~Balser, C.~Duelli, and W.~Reif.
\newblock Formal semantics of asbru - an overview.
\newblock {\em Proc. of the 6th Biennial World Conference on Integrated Design
  and Process Technology}, 5(5):1--8, 2002.

\bibitem{behrmann2004tutorial}
G.~Behrmann, A.~David, and K.~Larsen.
\newblock A tutorial on uppaal.
\newblock In {\em Formal Methods for the Design of Real-Time Systems}, pages
  200--236. Springer, 2004.

\bibitem{Christov2008Formally}
S.~Christov, B.~Chen, G.~S. Avrunin, et~al.
\newblock Formally defining medical processes.
\newblock {\em Methods of Information in Medicine}, 47(5):392, 2008.

\bibitem{Christov2008}
S.~Christov, B.~Chen, G.~S. Avrunin, et~al.
\newblock {\em Rigorously Defining and Analyzing Medical Processes: An
  Experience Report}, pages 118--131.
\newblock Springer Berlin Heidelberg, Berlin, Heidelberg, 2008.

\bibitem{fox1998disseminating}
J.~Fox, N.~Johns, and A.~Rahmanzadeh.
\newblock Disseminating medical knowledge: the proforma approach.
\newblock {\em Artificial Intelligence in Medicine}, 14(1-2):157 -- 182, 1998.

\bibitem{Guo2017COMPSAC}
C.~Guo, Z.~Fu, S.~Ren, Y.~Jiang, and L.~Sha.
\newblock Towards verifiable safe and correct medical best practice guideline
  systems.
\newblock In {\em 2017 IEEE 41st Annual Computer Software and Applications
  Conference (COMPSAC)}, July 2017.

\bibitem{Guo2016ICCPS}
C.~Guo, S.~Ren, Y.~Jiang, P.-L. Wu, L.~Sha, and R.~Berlin.
\newblock Transforming medical best practice guidelines to executable and
  verifiable statechart models.
\newblock In {\em 2016 ACM/IEEE 7th International Conference on Cyber-Physical
  Systems (ICCPS)}, pages 1--10, April 2016.

\bibitem{harel1987statecharts}
D.~Harel.
\newblock Statecharts: A visual formalism for complex systems.
\newblock {\em Science of computer programming}, 8(3):231--274, 1987.

\bibitem{yakindu}
Itemis.
\newblock Yakindu statechart tools (sct).
\newblock \url{https://www.itemis.com/en/yakindu/statechart-tools/}, July 2016.

\bibitem{StrokeGuideline}
E.~C. Jauch, B.~Cucchiara, O.~Adeoye, et~al.
\newblock Part 11: adult stroke: 2010 american heart association guidelines for
  cardiopulmonary resuscitation and emergency cardiovascular care.
\newblock {\em Circulation}, 122(18 suppl 3):S818--S828, 2010.

\bibitem{jiang2016use}
Y.~Jiang, H.~Liu, H.~Kong, R.~Wang, M.~Hosseini, J.~Sun, and L.~Sha.
\newblock Use runtime verification to improve the quality of medical care
  practice.
\newblock In {\em Software Engineering Companion (ICSE-C), IEEE/ACM
  International Conference on}, pages 112--121. IEEE, 2016.

\bibitem{jiang2017data}
Y.~Jiang, H.~Song, R.~Wang, M.~Gu, J.~Sun, and L.~Sha.
\newblock Data-centered runtime verification of wireless medical cyber-physical
  system.
\newblock {\em IEEE Transactions on Industrial Informatics}, 13(4):1900--1909,
  2017.

\bibitem{jiang2016stateflow}
Y.~Jiang, Y.~Yang, H.~Liu, H.~Kong, M.~Gu, J.~Sun, and L.~Sha.
\newblock From stateflow simulation to verified implementation: A verification
  approach and a real-time train controller design.
\newblock In {\em Real-Time and Embedded Technology and Applications Symposium
  (RTAS), 2016 IEEE}, pages 1--11. IEEE, 2016.

\bibitem{jiang2015design}
Y.~Jiang, H.~Zhang, Z.~Li, Y.~Deng, X.~Song, M.~Gu, and J.~Sun.
\newblock Design and optimization of multiclocked embedded systems using formal
  techniques.
\newblock {\em IEEE transactions on industrial electronics}, 62(2):1270--1278,
  2015.

\bibitem{Kim2010TII}
J.~Kim, I.~Kang, J.~Y. Choi, and I.~Lee.
\newblock Timed and resource-oriented statecharts for embedded software.
\newblock {\em IEEE Transactions on Industrial Informatics}, 6(4):568--578, Nov
  2010.

\bibitem{Mckinley2011computer}
B.~A. McKinley, L.~J. Moore, J.~F. Sucher, et~al.
\newblock Computer protocol facilitates evidence-based care of sepsis in the
  surgical intensive care unit.
\newblock {\em Journal of Trauma and Acute Care Surgery}, 70(5):1153--1167,
  2011.

\bibitem{McManus2016JCN}
M.~McManus and D.~S. Liebeskind.
\newblock Blood pressure in acute ischemic stroke.
\newblock {\em Journal of Clinical Neurology (Seoul, Korea)}, 12(2):137--146,
  2016.

\bibitem{patel1998representing}
V.~L. Patel, V.~G. Allen, J.~F. Arocha, and E.~H. Shortliffe.
\newblock Representing clinical guidelines in glif.
\newblock {\em Journal of the American Medical Informatics Association},
  5(5):467--483, 1998.

\bibitem{Prince2013strokeIA}
E.~A. Prince, S.~H. Ahn, and G.~M. Soares.
\newblock Intra-arterial stroke management.
\newblock {\em Seminars in interventional radiology}, 30(03):282--287, 2013.

\bibitem{Terenziani2004GLARE}
P.~Terenziani, S.~Montani, A.~Bottrighi, et~al.
\newblock The glare approach to clinical guidelines: main features.
\newblock {\em Stud. Health Technol. Inform.}, pages 162--166, 2004.

\bibitem{Tu2001EON}
S.~W. Tu and M.~A. Musen.
\newblock Modeling data and knowledge in the eon guideline architecture.
\newblock {\em Medinfo}, pages 280--284, 2001.

\bibitem{WuTreatment2014}
P.~Wu, D.~Raguraman, L.~Sha, R.~Berlin, and J.~Goldman.
\newblock A treatment validation protocol for cyber-physical-human medical
  systems.
\newblock In {\em Software Engineering and Advanced Applications (SEAA), 2014
  40th EUROMICRO Conference on}, pages 183--190, Aug 2014.

\bibitem{yang2016verifying}
Y.~Yang, Y.~Jiang, M.~Gu, and J.~Sun.
\newblock Verifying simulink stateflow model: timed automata approach.
\newblock In {\em Proceedings of the 31st IEEE/ACM International Conference on
  Automated Software Engineering}, pages 852--857. ACM, 2016.

\end{thebibliography}

\end{document}